\documentclass{article}
\usepackage{spconf,amsmath,graphicx}

\usepackage{caption} 
\usepackage{tabularx}
\usepackage{float}
\usepackage{array}  
\usepackage{booktabs} 
\usepackage{multirow}
\usepackage{makecell}

\usepackage{amsfonts,amssymb} 

\title{Rule-embedded network for audio-visual voice activity detection in live musical video streams}

\name{Yuanbo Hou$^1$, Yi Deng$^2$, Bilei Zhu$^3$, Zejun Ma$^3$ and Dick Botteldooren$^1$}
\address{$^1$Ghent University, Belgium \quad $^2$New York University, USA \quad $^3$Bytedance AI Lab, China.}
\begin{document}

\captionsetup{font={small}}

\maketitle
\begin{abstract}
Detecting \textit{anchor's} voice in live musical streams is an important preprocessing for music and speech signal processing. Existing approaches to voice activity detection (VAD) primarily rely on audio, however, audio-based VAD is difficult to effectively focus on the target voice in noisy environments. With the help of visual information, this paper proposes a rule-embedded network to fuse the audio-visual (\textit{A-V}) inputs to help the model better detect target voice. The core role of the rule in the model is to coordinate the relation between the bi-modal information and use visual representations as the mask to filter out the information of non-target sound. Experiments show that: 1) with the help of cross-modal fusion by the proposed rule, the detection result of \textit{A-V} branch outperforms that of audio branch; 2) the performance of bi-modal model far outperforms that of audio-only models, indicating that the incorporation of both audio and visual signals is highly beneficial for VAD. To attract more attention to the cross-modal music and audio signal processing, a new live musical video corpus with frame-level label is introduced.

\end{abstract}

\vspace{-0.1cm}
\section{Introduction}
\label{sec:intro}

With the rise of musical video social platforms such as TikTok and Douyin, more audio-visual (\textit{A-V}) data is uploaded to the Internet. To process these diverse data, VAD is an essential preprocessing to detect the presence or absence of human voice in clips. Classical applications of VAD include speaker diarization \cite{speaker_diarization}, music \cite{interspeech2020hyb} and speech \cite{speech_recognition} signal processing.

Many VAD methods \cite{lee2018revisiting} typically rely on acoustic features of audio signals \cite{hmm, lu2003automated}. These approaches perform well when there is little or no noise in background. Recently, deep learning brings more robust methods for VAD. A deep neural network using magnitude and phase information is proposed for VAD under noise condition \cite{dnnvad}. Convolutional neural networks have been used as acoustic models for VAD in noisy environments \cite{cnnvad}. An optimization process based recurrent neural networks for VAD is investigated in \cite{rnnvad}. 

In live musical streams, an \textit{anchor} is a person who speaks to the audience or sings in the front of the camera, and play other music at the same time. This paper aims to detect when the \textit{anchor} speaks or sings in live musical streams, which is a cocktail party-like environment where there may be more than one person singing or speaking at the same time, and the background contains music and other noise. Due to the interference of noise and voice from music in the background, it is easy to cause false detection and audio-based VAD may fail. That is, audio-based detectors are difficult to effectively detect target voice in live musical streams.

To make the detector better focus on the target voice and ignore unrelated sounds in the noisy background, a lot of attention has been paid to the \textit{A-V} VAD (AV-VAD). Considering the lip's movement is tightly correlated with speech, a statistical VAD model is designed to fuse the audio and lip's information \cite{liu2006audio}. By using face parameters (eye and lip contours), a real-time VAD is developed as a front-end for speech recognition \cite{2010real_vad}. \cite{diffusion} presents an \textit{A-V} detector to separate speech from non-speech frames. As mentioned above, previous works on AV-VAD mainly focus on speech. However, in live musical video streams, the \textit{anchor} usually spends more time singing than speaking, that is, the AV-VAD in this paper needs to detect not only speech, but also the singing of \textit{anchor}. Due to the different articulation and phonation between speaking and singing, the speech activity detector does not match well with musical clips \cite{interspeech2020hyb}. \cite{kosaka2018improving} attempts to detect singing voice and speech based on the same audio-based VAD model, without distinguishing whether the speech or singing voice comes from the \textit{anchor} or background. Currently, to our best knowledge, there is no direct research on singing voice detection using visual information in the diverse and noisy scenario like live musical streams. 

This paper aims to detect singing voice and speech of the \textit{anchor} in live musical streams, which is a mixture of polyphonic music and different voices from different people. To better concentrate on the target voice, a simple and effective rule-embedded network is proposed for AV-VAD based on a straightforward observation that visual events usually occur together with acoustic events and they are coordinated \cite{senocak2018learning}. The facial information can directly reflect whether the \textit{anchor} is vocalizing. This paper explores the possibility of using the facial information as the mask to filter out non-target acoustic events. Experiments show the incorporation of \textit{A-V} signals is highly beneficial for VAD in noisy scenarios. 

The main contributions of this work are: 1) the visual information is introduced into singing voice detection, and unlike previous audio-only VAD studies on either speech or singing voice, this work explores to detect speech and singing voice using \textit{A-V} information in live music streams; 2) to fuse the bi-modal inputs simply and reasonably, a rule-embedded network based on the instinctive observation is proposed; 3) to attract more attention to the cross-modal music and audio signal processing, a live musical video corpus is introduced.

This paper is organized as follows, Section 2 introduces the rule-embedded audio-visual multi-branch framework. Section 3 describes the dataset, baseline, experimental setup and analyzes the results. Section 4 gives conclusions.

\vspace{-0.1cm}
\section{Rule-embedded AV-VAD network}
\label{sec:format}

The proposed rule-embedded AV-VAD network is shown in Fig. \ref{model}, the left part is audio branch (red words) that tries to learn the high-level acoustic features of target events in audio level, and right part is image branch (blue words) attempts to judge whether the \textit{anchor} is vocalizing using visual information. The bottom part is the \textit{A-V} branch (purple italics), which aims to fuse the bi-modal representations to determine the probability of target events of this paper.

\subsection{The Audio Branch}

\vspace{-0.05cm}
The goal of audio branch is to learn core representations of 4 classes events: \textbf{Silence}, \textbf{Speech}, \textbf{Singing}, and \textbf{Others}. \textbf{Speech} and \textbf{Singing} contain all speech and singing voice in audio streams, without distinguishing whether the speech or singing voice comes from the anchor or background. The following \textit{A-V} branch will rely on visual information to determine the source of voices. \textbf{Others} class includes: non-speech and non-singing voice of the \textit{anchor}, such as coughing, laughing, cheering and instrumental music, etc.

\begin{figure}[tb]
	\setlength{\abovecaptionskip}{0.1cm}   
	\setlength{\belowcaptionskip}{-0.45cm}   
	\centerline{\includegraphics[width = 0.51  \textwidth]{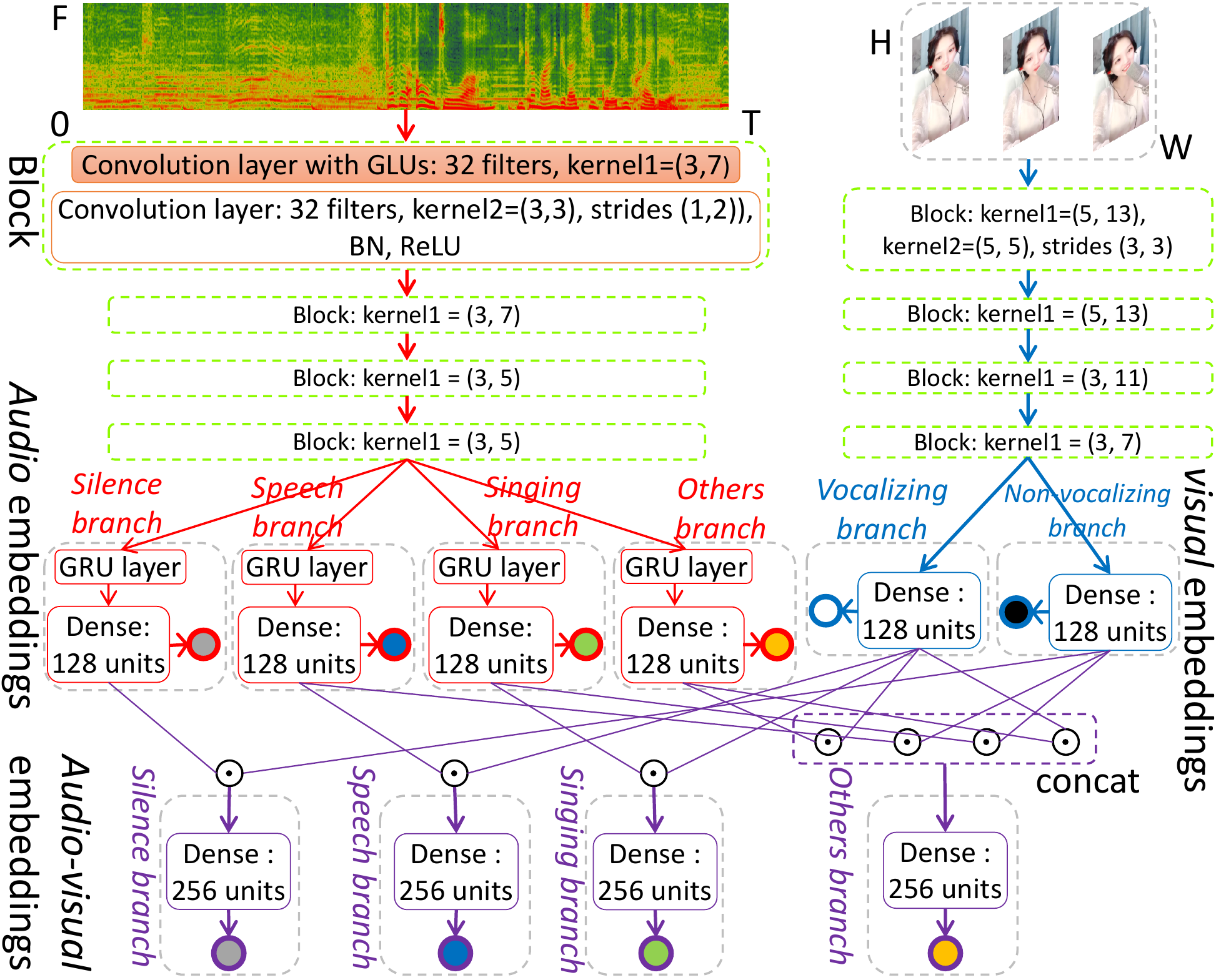}}
	\caption{The proposed rule-embedded AV-VAD network.}
	\label{model}
\end{figure}

To better learn the acoustic representations, a multi-output convolutional recurrent neural network (CRNN) is used, which is the audio branch connected by red lines in Fig. \ref{model}. Waveforms are converted to log mel spectrograms and input into the CRNN. Convolutional layers with gated linear units (GLUs) \cite{GLU} are applied to learn local shift-invariant patterns from the acoustic feature. By using GLUs, the model can learn to attend to target events and ignore unrelated sounds \cite{hou2019sound}. The pooling operation is implemented by convolution and strides is (1, 2), which means the stride of the convolution along the time is 1 to preserve the time resolution of the input.

 To learn the core representations of 4 classes events, 4 outputs with the same structure but different objectives are designed. After the last convolution layer, 4 gated recurrent units (GRU) \cite{gru} layers are adopted to capture the temporal context information of \textbf{Silence}, \textbf{Speech}, \textbf{Singing}, and \textbf{Others}, respectively. Following GRU layers, 4 dense layers are used to obtain the final latent representation of the 4 classes. The outputs of the audio branch are 4 independent units, each of them is a binary classification with sigmoid to represent the probability of the corresponding class in the current input. Readers could visit homepage\protect\footnote{https://github.com/Yuanbo2020/Audio-Visual-VAD} for details and source code.

\label{ssec:model}

\vspace{-0.15cm}
\subsection{The Image Branch}

The structure of image branch is similar to audio branch, but contains different goal. The image branch aims to predict the possibility of vocalizing and non-vocalizing of the \textit{anchor} using facial information in the image sequence. Hence, there are 2 outputs with the same structure to learn the core high-level representations of \textbf{Vocalizing} and \textbf{Non-vocalizing}.

To comprehensively consider the continuity of \textit{anchor's} actions between the front and rear frames and the contextual information of image in streams, the input of image branch is a fixed time length image sequence. A convolutional structure similar to that in the audio branch is used to extract spatial features of the image sequence. In experiments, we found that convolutional layers are more effective than recurrent layers in the image branch, so there is no GRU layer. Following convolutional layers, 2 dense layers are applied to extract high-level representations vectors. Like the audio branch, the outputs of image branch are 2 independent units with sigmoid function to indicate the probability of the corresponding class in the current input. This structure is to enable the model to learn the core embedding vector contained in each class.

\subsection{The Audio-Visual Branch}

Visual events and acoustic events are coordinated \cite{senocak2018learning}. The facial information can reflect whether the \textit{anchor} is vocalizing intuitively. When the \textit{anchor's} mouth is closed, it generally does not vocalize; when the \textit{anchor's} eyes are closed, it may be laughing, yawning, or sleeping instead of singing or speaking. Based on the correlation between the \textit{anchor's} voice activity and its face parameters, this paper explores the possibility of using the facial information as the mask to filter out non-target acoustic events and focus on the target voice event. The relation among various acoustic events in audio, visual events in image and the target events can be simply represented by Table \ref{figure1}. The main idea of Table \ref{figure1} is using visual representations as the mask to filter out the information of non-target sound. The relation in Table \ref{figure1} is simplified to the connection rule embedded in the network in Fig. \ref{model}. Note that \textbf{\textsl{Speech}} and \textbf{\textsl{Singing}} in \textit{A-V} branch just contain the speech and singing voice from \textit{anchor} (target speaker), and are different from the \textbf{Speech} and \textbf{Singing} in audio branch which may from background sounds. 

\label{ssec:figure1}
\begin{table}[b]
    \vspace{-0.5cm}  
	\caption{The relation among acoustic, visual and target events.}\label{figure1}
	\setlength{\abovecaptionskip}{-4cm}  
	\setlength{\belowcaptionskip}{-8cm} 
	\vspace{-0.3cm}
	\centerline{\includegraphics[width = 0.43  \textwidth]{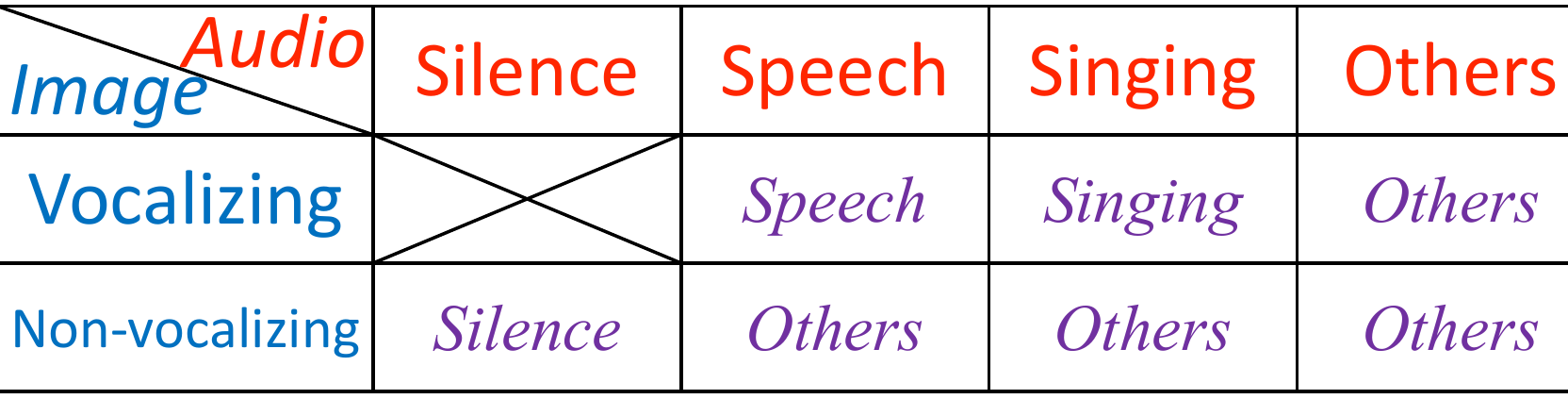}}
\end{table}

To simply and effectively combine the high-level representations of two branches with different semantics, this paper fuses the embedding vectors according to the rule in Table \ref{figure1}. And there are 3 common ways to fuse bi-modal vectors, including Hadmard product (HP) \cite{horn1990hadamard}, simple concat (SC) and matrix multiplication (MM). The dense layers in audio and image branch are used to obtain the final latent representation of each class, the corresponding vector is regarded as the abstract feature. Given acoustic and visual embeddings \{$\boldsymbol{a}, \boldsymbol{v}$\}${\in\mathbb{R}^{128}}$, the operations:

• SC: $\boldsymbol{o}$ $= Concat($$\boldsymbol{a}, \boldsymbol{v}$), $\boldsymbol{o}\in\mathbb{R}^{256}$;

• MM: $\boldsymbol{O}$ $= \boldsymbol{a}\boldsymbol{v}^\mathrm{T}$, $\boldsymbol{O}\in\mathbb{R}^{128\times128}$;

• HP: $\boldsymbol{o}$ $= \boldsymbol{a}\odot\boldsymbol{v}$, $\boldsymbol{o}\in\mathbb{R}^{128}$;

\noindent
where the symbol $\odot$ is the element-wise product. For unified processing, $\boldsymbol{o}$ and $\boldsymbol{O}$ will be flattened and input into the corresponding dense layer, as shown in Fig. \ref{model}. For bi-modal information, there is a simpler way than combining by the proposed rule, which is to directly look up table (LuT) according to Table \ref{figure1} based on the outputs of the audio and image branch. To comprehensively compare the performance of the proposed rule-embedded network, the performance of these methods will be shown in later experiments.

For training, to consider the cross-modal information at the same time and give the corresponding decisions, the audio, image and \textit{A-V} branch will be trained together. The loss function of the multi-branch AV-VAD model is:
\begin{equation}
	\setlength{\abovedisplayskip}{2.5pt}
	\setlength{\belowdisplayskip}{2.5pt}
	\begin{split}
	L=&\lambda_1L_{a-sil}+\lambda_2L_{a-spe}+\lambda_3L_{a-sin}+\lambda_4L_{a-oth}\\
	&+\lambda_5L_{v-voc} + \lambda_6L_{v-non-voc} + \lambda_7L_{av-sil} + \\
	&\lambda_8L_{av-spe} + \lambda_9L_{av-sin} + \lambda_{10}L_{av-oth} \\
	\end{split}
\end{equation}
where $L_a$, $L_v$ and $L_{av}$ denote the loss of audio, image and \textit{A-V} branch; $sil$, $spe$, $sin$ and $oth$ denote silence, speech, singing and others; $voc$ and ${non-voc}$ denote vocalizing and non-vocalizing. $\lambda_i$ is the scale factor of each loss function, $\lambda_i$ determines the importance of each loss function in training.

\vspace{-0.12cm}
\section{Experiments and results}
\vspace{-0.2cm}
\subsection{Dataset, Baseline, and Experiments Setup}

\vspace{-0.1cm}
Since there is no direct research on singing voice detection by \textit{A-V} information in live streams, and there are fewer \textit{A-V} datasets that can be used to train the AV-VAD model to detect both target speech and singing voice in noisy environments, a 500-minute musical video dataset (MVD500) with frame-level label is used in this paper. In detail, 360 minutes for training, 40 minutes for validation and 100 minutes for testing. To avoid bias caused by gender in training, the total duration of live broadcast of male and female \textit{anchors} is close in MVD500. And to attract more attention to the cross-modal musical signal processing, a 100-minute musical \textit{A-V} corpus (MAVC100) with frame-level label is introduced in our homepage for open research. 
 
For baselines, a common and typical bi-modal recurrent neural model \cite{tao2019end} is used as \textit{A-V} baseline (\textsl{Base-AV}), and a CRNN \cite{interspeech2020hyb} trained by transfer learning is used as audio-based baseline (\textsl{Base-A}) to compare the performance of the AV-VAD from more perspectives.

In training, log mel-band energy is extracted using STFT with Hamming window length of 44 ms and the overlap is 50\% between the window, which has sufficient time-frequency resolution. Then 64 mel filter banks are used. Dropout and normalization are used to prevent over-fitting. Each output in the AV-VAD model is binary classification, Adam optimizer \cite{adam} is used to minimize the binary cross entropy. Please see our homepage for source code.

For evaluation metrics, event-based precision (\textsl{P}), recall (\textsl{R}), \textsl{F-score} and Error rate (\textsl{ER}) \cite{metrics} are used. Compared with segment-based metrics used in previous studies \cite{Mesaros2019_TASLP, hou2019sound, interspeech2020hyb}, event-based metrics are more rigorous and accurate to measure the location of events. Higher \textsl{P}, \textsl{R}, \textsl{F} and lower \textsl{ER} indicate a better performance.
 
\vspace{-0.3cm}
\subsection{Results and Analysis}

\vspace{-0.1cm}
This section tries to answer the following \textbf{r}esearch \textbf{q}uestions:
\textbf{• RQ1}: Which cross-modal fusion way is the most effective? Based on the same bi-modal input and combined by rule, how about the performance of the fusion by HP, SC, MM and LuT?

\vspace{-0.03cm}
\begin{table}[b]\footnotesize
	\setlength{\abovecaptionskip}{0cm}   
	\setlength{\belowcaptionskip}{-0.55cm}   
	\renewcommand\tabcolsep{1.5pt} 
	\centering
	\caption{The detection results on the test set of MVD500.}
	\begin{tabular}{p{1.7cm}<{\centering}p{1.0cm}<{\centering}p{1cm}<{\centering}p{1cm}<{\centering}p{1cm}<{\centering}p{1.4cm}<{\centering}} 
		\toprule[1pt] 
		\specialrule{0em}{0.1pt}{0.1pt}
		& \textsl{SC} & \textsl{MM} & \textsl{HP} & \textsl{LuT} & \textsl{Base-AV}\\
		\midrule[1pt]  
		\specialrule{0em}{0.1pt}{0.1pt}
		\textsl{ER} & 0.53 & 0.48 & \textbf{0.40} & 0.78 & 0.72\\
		\specialrule{0em}{0.1pt}{0.1pt}
		\textsl{P (\%)} & 75.19 & 78.97 & \textbf{84.95} & 62.70 & 66.17\\
		\specialrule{0em}{0.2pt}{0.2pt}
		\textsl{R (\%)} & 66.54 & 67.43 & \textbf{70.77} & 49.25 & 53.41\\
		\specialrule{0em}{0.2pt}{0.2pt}
		\textsl{F-score (\%)} & 70.60 & 72.75 & \textbf{77.22} & 55.17 & 59.11\\
		\specialrule{0em}{0.2pt}{0.2pt}
		\bottomrule[1pt]
		\specialrule{0em}{0pt}{0pt}
	\end{tabular}
	\label{tab:rule}
\end{table}

To demonstrate the effect of these fusion methods in the cross-modal learning, the detection results of different fusion ways on the test set of MVD500 are shown in Table \ref{tab:rule}. Since the activation function of last classification layer in \textsl{Base-AV} is Softmax, which means the output units are equally important in \textsl{Base-AV}. So, for a fair comparison between the multi-branch model in this paper and \textsl{Base-AV}, all $\lambda_i$ in $L$ are 1.

In Table \ref{tab:rule}, the cross-model fusion by HP has the best effect, followed by MM, and SC has the worst effect. The reason may be that HP is a binary operation, based on HP, the visual embedding vector functions as a mask to help the model attend to target voice and ignore unrelated sound in the noisy background. Each bit in the high-level acoustic feature vector is adjusted via the corresponding visual information, the audio representations that are beneficial to the task are retained and irrelevant content is reduced. SC retains the information without any filtering or enhancement effect, and does not guide the exploration of model. The output of MM has richer information such as the correlation between different bits of audio and image vector, but these information are not independent of each other. Because $rank(\boldsymbol{O})$ is 1, the output matrix of MM is rank deficient. And after flattening, the 128$\times$128-dimensional output of MM bring a great burden to the calculation of the model. 

All results of rule-based combination (SC, MM and HP) are better than the \textsl{Base-AV}. This may be because the audio and image branch of \textsl{Base-AV} are just a common feature extractor, and not optimized for the task. The network structure in this paper is specifically designed for AV-VAD, so that it can learn the specific high-level feature of target event, rather than the universal representation of all events. The result of LuT also show the two branches in AV-VAD network is effective, because LuT does not involve cross-modal fusion, and only combines the results of two branches according to Table \ref{figure1}. The result of LuT is not much worse than that of \textsl{Base-AV}, indicating that the output of two branches did learn the knowledge of corresponding events as we expected.

\noindent
\textbf{• RQ2}: How much improvement is the result of \textit{A-V} branch compared with that of audio branch in the same bi-modal model? That is, if the bi-modal inputs and combination rule are useful, using visual information as the mask to filter out irrelevant acoustic representations in the audio, how much gain can this operation bring to the final result?

\begin{table}[b]\small  
	\setlength{\abovecaptionskip}{0.2cm}   
	\setlength{\belowcaptionskip}{0cm}   
	\renewcommand\tabcolsep{0.5pt} 
	\centering
	\vspace{-0.4cm}
	\caption{The gain of visual information.}
	\begin{tabular}
	{ p{1.7cm}<{\centering}|
	  p{1.8cm}<{\centering}|
	  p{1cm}<{\centering}
	  p{1cm}<{\centering}
	  p{1cm}<{\centering}
	  p{1.7cm}<{\centering}} 
		\hline
		\multicolumn{2}{c|}{\textbf{\textsl{Results of branch}}} & \textsl{ER} & \textsl{P (\%)} & \textsl{R (\%)} & \textsl{F-score (\%)}\\
		\hline
		\multirow{2}{*}{\makecell[c]{\textbf{\textsl{AV-VAD}}}}  & \textsl{audio} & 0.56 & 70.94 & 68.10 & 69.49 \\
		\cline{2-6}
		 & \textsl{audio-visual} & 0.40 & 84.95 & 70.77 & 77.22\\
		\hline
		\multirow{2}{*}{\makecell[c]{\textbf{\textsl{Base-AV}}}}  & \textsl{audio} & 0.75 & 64.95 & 50.50 & 56.88\\
		\cline{2-6}
		 & \textsl{audio-visual} & 0.72 & 66.17 & 53.41 & 59.11\\
		\hline
	\end{tabular}
	\label{tab:image}
\end{table}

The gain of visual information in the cross-modal model is shown in Table \ref{tab:image}, both AV-VAD and \textsl{Base-AV} achieve the effect of the output of \textit{A-V} branch is better than that of audio branch, which shows both models coordinate the relation between the \textit{A-V} information well, so that the result of the fusion of different modal information outperforms that of single modal information. In detail, the gain of AV-VAD on the \textsl{F-score} and \textsl{ER} is 7.73\% and 0.16, while that of \textsl{Base-AV} is 2.23\% and 0.03. AV-VAD is not only on the audio branch, but its corresponding visual information gain is also better than the \textsl{Base-AV}.

\noindent
\textbf{• RQ3}: Compared with audio-based VAD, how much improvement is the performance of AV-VAD?

\begin{figure}[t]
	\vspace{-0cm}  
	\setlength{\abovecaptionskip}{0.2cm}   
	\setlength{\belowcaptionskip}{-0.3cm}   
	\centerline{\includegraphics[width = 0.5  \textwidth]{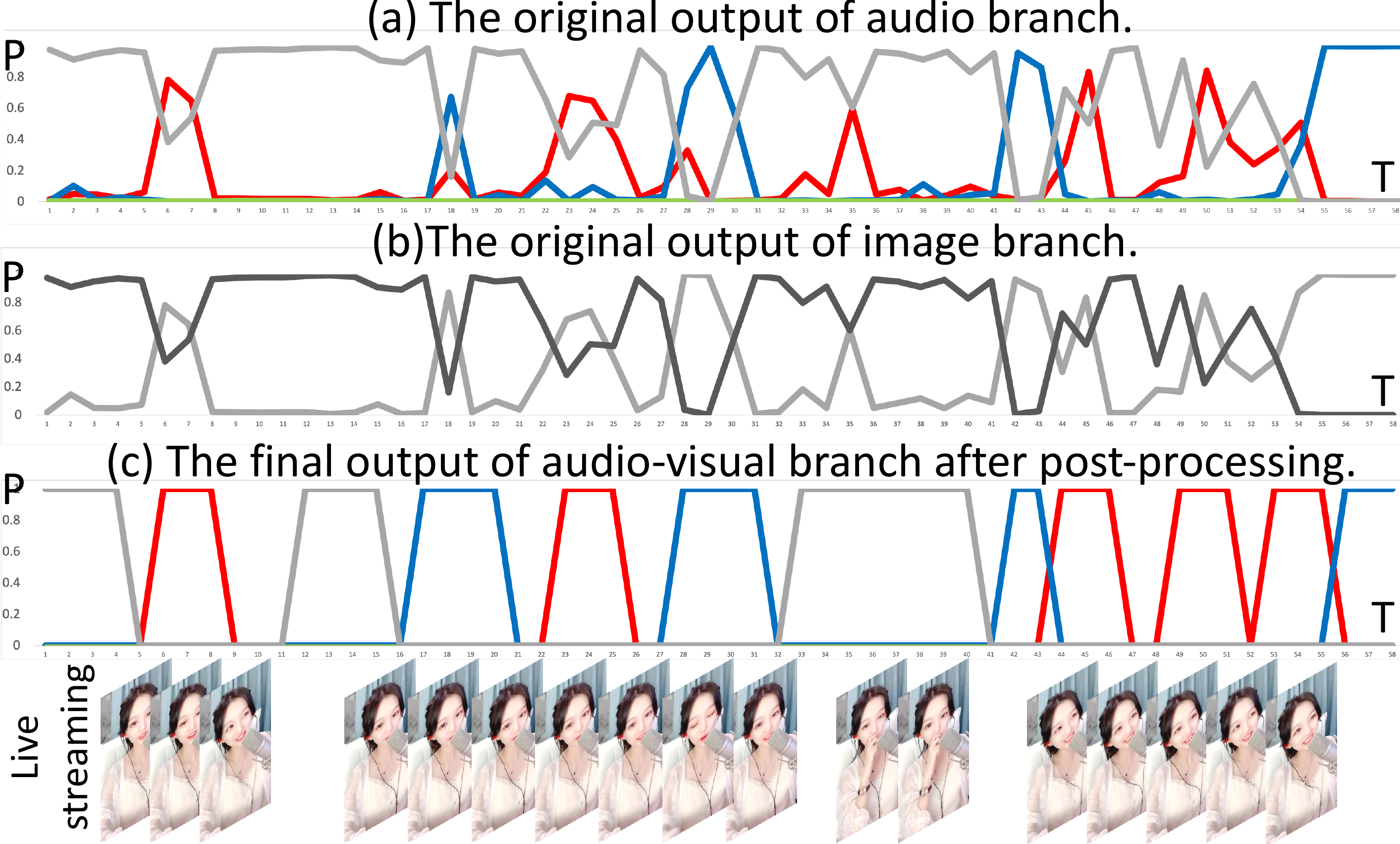}}
	\caption{In subgraph (a), the red, blue, gray and green lines denote \textbf{Singing}, \textbf{Speech}, \textbf{Others} and \textbf{Silence} in audio. In subgraph (b), the gray and black lines denote vocalizing and non-vocalizing. In subgraph (c), the red, blue and gray lines denote the target \textbf{\textsl{Singing}}, \textbf{\textsl{Speech}} and \textbf{\textsl{Others}}, and the other remaining part is \textbf{\textsl{Silence}}.}
	\label{figure-f}
\end{figure}

The \textsl{ER} and \textsl{F-score} of the audio-based \textsl{Base-A} is 0.86 and 40.93\%, respectively. And the corresponding values of AV-VAD shown in Table \ref{tab:image} are 0.40 and 77.22\%, respectively. The performance of bi-modal model far outperforms that of audio-based \textsl{Base-A}, which means the rule-embedded AV-VAD is more effective in noisy audio-visual environment. A intuitive detection result of AV-VAD from the live broadcast is shown in Fig. \ref{figure-f}. Please see our homepage for video demos.

\label{ssec:figure-f}

The above experiments are all tested on the test set of MVD500. To make it easier for others to compare our method and show results more comprehensively, the detection results of AV-VAD on public MAVC100 is shown in Table \ref{tab:MAVC100}.

\begin{table}[H]\small  
	\setlength{\abovecaptionskip}{0.2cm}   
	\setlength{\belowcaptionskip}{0cm}   
	\renewcommand\tabcolsep{0.5pt} 
	\centering
	\caption{Main results of AV-VAD on the public MAVC100.}
	\begin{tabular}
	{ p{1.2cm}<{\centering}|
	  p{0.85cm}<{\centering}
	  p{0.9cm}<{\centering}
	  p{0.85cm}<{\centering}
	  p{0.9cm}<{\centering}|
	  p{0.85cm}<{\centering}
	  p{0.9cm}<{\centering}
	  p{0.85cm}<{\centering}
	  p{0.9cm}<{\centering}} 
	  
		\hline
		
		\multirow{2}{*}{\makecell[c]{\textbf{\textsl{Class}}}}& \multicolumn{4}{c|}{\textbf{\textsl{Event-based}}} & \multicolumn{4}{c}{\textbf{\textsl{Segment-based}}}\\
		
		\cline{2-9}
		
		 & \textsl{ER} & \textsl{P (\%)} & \textsl{R (\%)} & \textsl{F (\%)} & \textsl{ER} & \textsl{P (\%)} & \textsl{R (\%)} & \textsl{F (\%)}\\
		\hline
	
		\textsl{Singing} & 0.48 & 38.36 & 49.59 & 70.11 & 0.13 & 92.65 & 93.57 & 93.11\\
		
		\hline
		\textsl{Speech} & 0.36 & 90.65 & 70.78 & 79.49 & 0.21 & 97.01 & 80.85 & 88.20\\
		
		\hline
		\textbf{\textsl{Overall}} & 0.44 & 80.83 & 70.40 & 75.26 & 0.15 & 95.35 & 86.93 & 90.94 \\
		\hline
		
	\end{tabular}
	\label{tab:MAVC100}
\end{table}

\section{CONCLUSION}
\label{sec:CONCLUSION}

This paper proposes a rule-embedded network to fuse the cross-modal information for audio-visual VAD in live musical video streams. Experiments show the proposed cross-modal fusion method and multi-branch network are effective, and the performance of bi-modal model far outperforms that of audio-only model. Besides, to attract more attention to the multi-modal music and audio signal processing, a new live musical video corpus with frame-level label is introduced.

\setlength{\parindent}{2.0em}

\vfill\pagebreak

\label{sec:refs}

\bibliographystyle{IEEEbib}
\bibliography{Template}

\end{document}